\documentclass[12pt]{iopart}
\pdfoutput=1 
\usepackage{iopams}
\usepackage{graphicx}
\usepackage{amssymb}
\usepackage{amsfonts}
\usepackage{bm}
\usepackage[utf8]{inputenc}
\usepackage[T1]{fontenc}
\usepackage{color}
\usepackage[normalem]{ulem}
\usepackage{subfigure}

\newcommand{\bn}[1]{\mbox{\boldmath $#1$}}

\newcommand{\mb}{\mbox}

\begin{document}


\title[Permutations of the effective valence-band {potential} for layered heterostructures.]{Permutations of the transverse momentum dependent effective valence-band {potential }for layered heterostructures. Pseudomorphic strain effects.}

\author{L. Diago-Cisneros$^{1,2}$ J. J. Flores-Godoy$^1$ A. Mendoza-\'{A}lvarez$^1$, and G. Fern\'{a}ndez-Anaya$^1$}
\ead{ldiago@fisica.uh.cu}

\address{$^1$Departamento de F\'{\i}sica y Matem\'{a}ticas, Universidad Iberoamericana,  M\'exico D. F., C.~P. 01219, M\'{e}xico.}
\address{$^2$ Facultad de F\'{\i}sica, Universidad de La Habana C.~P. 10400, Cuba.}
\date{\today}


\date{\today}

\begin{abstract}
The evolution of transverse-momentum-dependent effective band offset ($V_{\mathrm{eff}}$) profile for heavy (\emph{hh})- and light-holes (\emph{lh}), is detailed studied. Several new features in the metamorphosis of the standardized {fixed-height} $V_{\mathrm{eff}}$ profile {for holes}, in the presence of gradually increasing valence-band mixing {and pseudomorphic strain}, are presented. In some  $III-V$ unstrained semiconducting layered heterostructures a {fixed-height} potential, {is not longer valid for \emph{lh}. Indeed, we found --as predicted for electrons--, permutations of the $V_{\mathrm{eff}}$ character for \emph{lh}, that resemble a ``\emph{keyboard}", together with bandgap changes}, whenever the valence-band mixing varies from low to large intensity. 
{Strain is able to diminish the \emph{keyboard} effect on $V_{\mathrm{eff}}$, and also makes it emerge or vanish occasionally.} {We found that}  multiband-mixing effects and stress induced events, {are competitors mechanisms} that can not be universally neglected by assuming a { fixed-height rectangular spatial} distribution for {fixed-character} potential energy,  {as a reliable test-run input for heterostructures}. Prior to the present report, neither direct transport-domain measurements, nor theoretical calculations addressed to these $V_{\mathrm{eff}}$  {evolutions and permutations}, has been reported for holes, as far as we know. {Our results may be of relevance for promising heterostructure's design guided by valence-band structure modeling to enhance the hole mobility in $III-V$ materials.}
\end{abstract}

\pacs{71.70.Ej, 72.25.Dc, 73.21.Hb, 73.23.Ad}
\maketitle

\section{Sinopsis of Fundamentals and Motivation}
\label{sec:Funda}

\hspace{5mm}For many actual practical solutions and technological applications, due to the impressive development of low-dimensional electronic and optoelectronic devices, it is drastically important to include the valence-band mixing \cite{Klicmeck01}, \textit{i.e.} the degree of freedom transverse to the main transport direction, whenever the holes are involved. This phenomenology, early quoted by Wessel and Altarelli in resonant tunneling \cite{Wessel89}, has been lately pointed up for real-life technological devices \cite{Klicmeck01}. If the electronic transport through these systems, engage both electrons and holes, the low-dimensional device response depends on the slower-heavier charge-carrier's motion through specific potential regions \cite{Schneider89}. It is unavoidable to recognize that in the specialized literature there is plenty of reports studying several physical phenomena derived from hole mixing effects and  {strain}, \textit{via} standard existing methods.  {Some authors had managed to determine optimal situation in a resonant tunneling of holes under internal strains, disregarding scattering effects and assuming a spatial symmetry for a constant potential} \cite{Bittencourt97}.  {A fundamental study on valence-band mixing in first-principles, established a non-linear response for a pseudo-potential in series of the atomic distribution function} \cite{Foreman07}.  {Valence-band mixing and/or strain had been extensively studied over the past few decades in several nanostructures ranging from quantum wells \cite{Bittencourt97,Ekins99,Smeeton03,Nainani11}, to quantum wires \cite{Faux97,Sunil08} and to quantum dots \cite{Bafna13}}. However, just a few reports are available, concerning the very evolution itself of the effective potential due to several causes, as a central topic of research.   {We underline in the present paper the focus not on valence-band mixing and strain effects problem in general, but rather on the particular metamorphosis of the effective potential while manipulating both  effects. We hope to make some progress in understanding the underlaying physics as well as determine whether or not the valence-band mixing and strain are competitors mechanism in the evolution of the effective potential}.

Earliest striking elucidations due to Milanovi\'{c} and Tjapkin for electrons,\cite{Milanovic82} and recalled much later by P\'erez-\'Alvarez and Garc\'{\i}a-Moliner for a fully unspecific multiband theoretical case,\cite{RF04} are fundamental cornerstones in this concern. The metamorphosis of the effective \emph{band offset} potential $V_{\mathrm{eff}}$, ``felt" by charge carriers depending on the transverse momentum value, is so far, the better way to graphically mimic, the phenomenon of the in-plane dependence of the effective mass, also refereed as the valence-band mixing for holes. In few words, a hole band mixing is crucial for bulk and low-dimensional confined systems possessing quantal heterogeneity, a question soon to be discussed in this paper, inspired in a similar scenario, as was done before for a single-band-electron problem \cite{Milanovic82}. Particularities of the appealing evolution features of $V_{\mathrm{eff}}$ for holes, in the presence of gradually increasing valence-band mixing  {and strain}, could be of interest for condensed-matter physicists, working in the area of quantum transport for multiband-multichannel models. While previous studies in quantum systems have added substantial contributions to the elucidation of the valence-subband structure \cite{Klicmeck01}, and the influence of the hole mixing on it \cite{Foreman07,Milanovic82,RF04}, there remain some aspects which do not appear to have received yet sufficient attention and/or because of their interest deserve further clarification. This is essentially the case of  {the strain influence together with} carriers' transverse motion connection with the $V_{\mathrm{eff}}$ they interact with,  {which are} the main porpoises under investigation here. We assume the last, widely understood as crucial for charge and spin carriers' quantum transport calculations through standard quantum barrier(QB) -- quantum well(QW) layered systems.

On general grounds, $V_{\mathrm{eff}}$ is given by the difference for $3D$ band edge levels  as long as the transverse momentum ($\kappa_{\mb{\tiny T}}$) values are negligible.\cite{RF04} For finite $\kappa_{\mb{\tiny T}}$, this assertion is no longer valid and the mixing effects reveal. The mechanism responsible for this behavior, is the increment of the $\kappa_{\mb{\tiny T}}$-quadratic proportional term, yielding even to invert the roles of QW and QB \cite{Milanovic82,RF04}. Some authors had declared a shift upward in energy, of the bound states in the effective potential well as the transverse wave vector increases \cite{Ekbote99}. By letting grow $\kappa_{\mb{\tiny T}}$, were found the valence-band mixing effects to arise and $V_{\mathrm{eff}}$ to change \cite{LCRP06}. They conclude a larger reduction for $V_{\mathrm{eff}}$ as a function of $\kappa_{\mb{\tiny T}}$, for light holes (\emph{lh}) respect to that for heavy holes (\emph{hh}) \cite{LCRP06}. These former works \cite{Foreman07,Milanovic82,RF04,Ekbote99,LCRP06}, were motivating enough and put us on an effort to try a more comprehensive vision, of how $V_{\mathrm{eff}}$ evolves spatially with $\kappa_{\mb{\tiny T}}$ {and strain}, for \emph{hh} and \emph{lh}. This paper is devoted to demonstrate, the feasibility of  {the } $V_{\mathrm{eff}}$ profile  {evolution},  {QB-QW permutations, and bandgap changes}, as a reliable follow-up  {tools} for finding the response of a layered semiconductor system ---with spatial-dependent effective mass---, on travelling holes throughout it, by tuning the valence-band mixing and including stress effects.

\begin{figure}
 \centering
 \includegraphics[width=5in,height=2.8in]{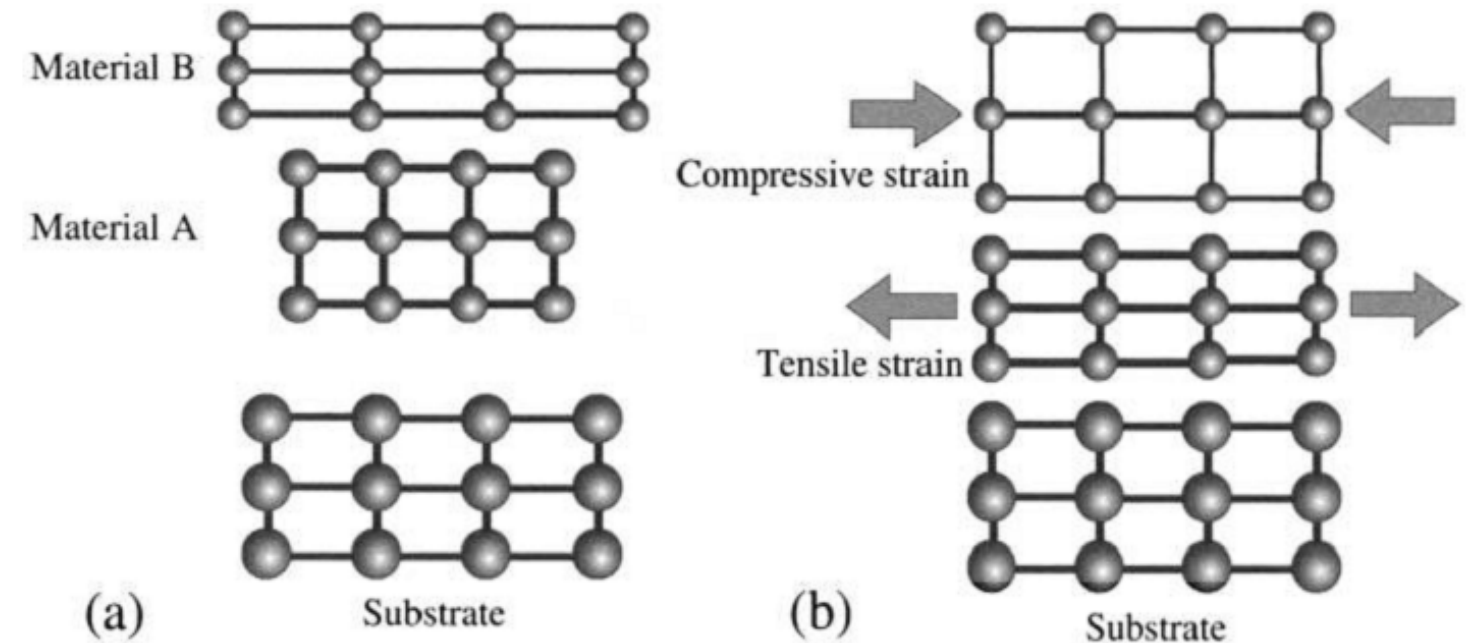}
   \caption{\label{F1-2Dn}  Panel (a) shows the stress-free bulk materials, with lattice parameter $a_{l} < a_{s}$ smaller (\textsl{GaAsP}), and larger $a_{l} > a_{s}$ (\textsl{InGaAs}) than that of the substrate.\cite{Piprek03} Panel (b) illustrates a schematic representation of a pseudomorphic grown process for a layered heterostructure.\cite{Jovanovic05} The material \textsl{GaAsP} is under a tensile strain, while the material (\textsl{InGaAs}) is under compressive strain, as they both are forced to conform the buffer's lattice constant $a_{s}$ of a suitable semiconductor wafer.}
\end{figure}

Built-in elastic strained layered heterostructures, has been remarkably used in the last decade, for development of light-emitting diodes, lasers, solar cells and photodetectors.\cite{RMohan10} Besides, internal strain may results into a considerably modification of the electronic structure of both electrons and holes, thereby altering the response of strained systems respect to nominal behavior of strain-free designs.\cite{RMohan10} We get motivated about the probable existence of a competitor mechanism able to diminish the effects of valence-band mixing on $V_{\mathrm{eff}}$, or wipe them out occasionally. Thus, owing to the need to account for strain in the present study, we additionally suppose the heterostructure sandwiched into an arbitrary configuration of pseudomorphically strained sequence of QW-acting and QB-acting binary(ternary) allows, due technological interest in that configuration. Whenever on layer-by-layer deposition, the epitaxially grown layer's lattice parameter matches that of the substrate in the in-plane direction --without collateral dislocations or vacancies--, the process is referred as pseudomorphic [see Fig. \ref{F1-2Dn}(b)].\cite{Jovanovic05} The last is widely chosen for most day-to-day applications, designed on a write-read platform, such as sound/image players and data-manipulating devices.

The outline for this paper is the following: Section \ref{sec:PotEff} presents briefly the theoretical framework to quote valence-band $V_{\mathrm{eff}}$ for both unstressed and stressed systems. Graphical simulations on $V_{\mathrm{eff}}$ evolution, are exposed in Section \ref{sec:Res&Dis}. In that section, we exercise and discuss highly specialized $III-V$ semiconductor binary(ternary)-compound cases, that support the main contribution of the present study  {and suggest possible applications}. Section \ref{sec:Conclu}, contains some conclusions.

\section{Calculation of the effective potential}
\label{sec:PotEff}

\hspace{5mm}Commonly, a wide class of solid-state physics problems, related to electronic and transport properties, demands the solution of multiband-coupled differential system of equations, widely known as Sturm-Liouville
matrix generalized boundary problem \cite{RF04}:
 \begin{eqnarray}
  \label{Eqmaestra}
   \frac{d}{dz} \left[ \bn{B}(z) \frac{d\bn{F}(z)}{dz} + \bn{P}(z)
   \bn{F}(z)\right] + \bn{Y}(z) \frac{d\bn{F}(z)}{dz} + \bn{W}(z) \bn{F}(z)
   & = & \bn{O}_{\mb{\tiny N}},
 \end{eqnarray}
\noindent where $\bn{B}(z)$ and  $\bn{W}(z)$ are, in general, $(N \times N)$ Hermitian matrices and is fulfilled $\bn{Y}(z) = -\bn{P}^{\dagger}(z)$. In the absence of external fields, standard plane-wave solutions are assumed and it is straightforward to derive a non-linear algebraic problem
\begin{equation}
 \label{eq:QEP}
 \bn{Q}(k_{z})\bn{\Gamma} = \left\{k_{z}^{2}\bn{\mathcal{M}} + k_{z}\,\bn{\mathcal{C}} + \bn{\mathcal{K}} \right\}{\mb{\bn{\Gamma}}}
                  = \bn{O}_{\mb{\tiny N}},
\end{equation}
\noindent called as quadratic eigenvalue problem (QEP),\cite{LCRP06} since  $\bn{Q}(k_{z})$ is a second-degree matrix polynomial on the $z$-component wavevector $k_{z}$. In the specific case of the well-known ($4 \times 4$) Kohn-Lüttinger (KL) model Hamiltonian, the matrix coefficients of equation (\ref{eq:QEP}) bear a simple relation with those in (\ref{Eqmaestra}) \cite{LCRP06}:
\begin{equation}
 \label{QEP-Eqmaestra}
  \bn{\mathcal{M}} = -\bn{B},\;\; \bn{\mathcal{C}} = 2i\bn{P} \;\;\mbox{and}\;\; \bn{\mathcal{K}} = \bn{W}.
\end{equation}
Then for ($4 \times 4$) KL model, the matrix coefficients of (\ref{eq:QEP}) can be cast as :
\begin{equation}
 \label{M-KL}
 \bn{\mathcal{M}} =
 \left(
  \begin{array}{cccc}
    -m^{*}_{hh} & 0 & 0 & 0 \\
    0 & -m^{*}_{lh} & 0 & 0 \\
    0 & 0 & -m^{*}_{h} & 0 \\
    0 & 0 & 0 & -m^{*}_{hh}
  \end{array}
 \right)
\end{equation}
 \begin{equation}
  \bn{\mathcal{C}} =
   \left(
    \begin{array}{cccc}
    0 & 0 & h_{13}+iH_{13} & 0 \\
    0 & 0 & 0 & -h_{13}-iH_{13} \\
    h_{13}-iH_{13} & 0 & 0 & 0 \\
    0 & -h_{13}+iH_{13} & 0 & 0
   \end{array}
  \right)
 \end{equation}
\begin{equation}
 \bn{\mathcal{K}} =
 \left(
  \begin{array}{cccc}
    a_{1} & h_{12}+iH_{12} & 0 & 0 \\
    h_{12}-iH_{12} & a_{2} & 0 & 0 \\
    0 & 0 & a_{2} & h_{12}+iH_{12} \\
    0 & 0 & h_{12}-iH_{12} & a_{1}
  \end{array}
 \right)
\end{equation}
\noindent Here $m^{*}_{hh,lh}$ stands for the (\textit{hh},\textit{lh}) effective mass, respectively. We briefly introduce some parameters and relevant quantities (in atomic units) of the KL model:

$\gamma_{i}$, with $i=1,2,3$ [Lüttinger semi-empirical valence band parameters, typical for each semiconductor material].

$R=13.60569172$ eV [Rhydberg constant],

$a_{0}=0.5405$ {\AA} [Bohr radius],

$V$ [Finite stationary barrier's height].

$E$ [Energy of incident and uncoupled propagating modes],

$k_{x}$, $k_{y}$ [Components of the transversal wavevector],

$A_{1,2}=a_{0}^{2}R\left(\gamma_{1} \pm \gamma_{2}\right)$,

$a_{1,2}=A_{1,2}\kappa_{\mb{\tiny T}}^{2} + V(z) - E$,

$h_{12}=a_{0}^{2}R\sqrt[2]{3}\gamma_{2}\left(k_{y}^{2} - k_{x}^{2}\right)$,

$h_{13}=-a_{0}^{2}R\sqrt[2]{3}\gamma_{3}k_{x}$,

$H_{13} = a_{0}^{2}R\sqrt[2]{3}\gamma_{3}k_{y}$,

$H_{12}=a_{0}^{2}R\sqrt[2]{3}2\gamma_{3}k_{x}k_{y}$,

Bearing direct association to the original matrix dynamic equation, we exclusively focus to the case when $\bn{\mathcal{M}}$, $\bn{\mathcal{C}}$ and $\bn{\mathcal{K}}$ are constant-by-layer, hermitian; and  $\bn{\mathcal{M}}$ is non-singular; therefore $k_{z}$ are all different real (symmetric) or arises in conjugated pairs $(k_{z},k_{z}^{*})$. Hereafter $\bn{O}_{\mb{\tiny N}}/\bn{I}_{\mb{\tiny N}}$, stand for $(N \times N)$ null/identity matrix. The QEP's solutions result in the eigenvalues $k_{z_{j}}$ and the eigenvectors ${\mb{\bn{\Gamma}}_j}$. As $\bn{Q}(k_{z})$ is regular, eight finite-real or complex-conjugated pairs of eigenvalues are expected.  Assuming a QEP method,\cite{LCRP06,AJGL11} it can be cast
\begin{equation}
 \label{eq:Polinom}
 \det\left[\bn{Q}(k_{z})\right] = q_{0} k_{z}^{8} + q_{1} k_{z}^{6} + q_{2} k_{z}^{4} + q_{3} k_{z}^{2} + q_{4},
\end{equation}
\noindent which is an eighth-degree polynomial with only even power of $k_z$ and real coefficients. The coefficients $q_{i}$ are functions of the system's parameters, and $q_{0} = \det\bn{\mathcal{M}}$ as expected.\cite{AJGL11} In the specific case of the Kohn-Lüttinger (KL) model Hamiltonian,\cite{LCRP06} $q_{i}$ contain the L\"uttinger semi-empirical valance-band parameters and the components of in-plane quasi-wave vector $\kappa_{\textsc t}$.

Based on our \textmd{procedure} \cite{AJGL11}, it is straightforward to follow whereas $k_{z}$ is oscillatory or not  {by dealing with} (\ref{eq:Polinom}), and thereby the kind of $V_{\mathrm{eff}}$ the holes interplay with. We will refer to \emph{root-locus-like} terminology from now on throughout the paper, whenever we proceed to invoke a complex-plane dependence, for the QEP (\ref{eq:QEP}) eigenvalues evolution as the valence-band mixing parameter changes. To our knowledge, just few pure theoretical or numerical applications of the \emph{root-locus-like} technique, particularly for the QEP scenario, had been previously addressed to explicitly describe several standard $III-V$ semiconductor compounds \cite{AJGL11,JALG13}. We get motivated  {by the advantages of the \emph{root-locus-like} technique in solid state physics \cite{AJGL11,JALG13}, and try to foretell here}, new features of the particle-scatterer interaction  {in the presence of valence-band mixing and strain}. For some high specialized zinc-blenda and wurtzite systems, current knowledge of the  {hole} quantum transport mechanism, is far to be profound. The present theoretical contribution, claim to spread light on that issue. Mainly, we think here in readers that may be interested more on the way the effective valence-band offset metamorphosis with band mixing and strain, could influence on their day-to-day applications, rather than getting involved with the very details of the theoretical model itself, only. Owing to that concern, we propose a simple and comprehensive modelling procedure for $V_{\mathrm{eff}}$ to deal with, and a \textit{gedanken}-like simulation for a passage of mixed holes throughout strained-free and strained layered heterostructures is exercised.

An effective potential, is found useful to describe valence-band mixing in the EFA.\cite{Foreman07} In the case envisioned here, to determine the operator $\widehat{\bn{W}}_{\mathrm{eff}}$ for the effective \emph{band offset} potential, suffices to use the Kohn-L\"uttinger (KL) model Hamiltonian,\cite{LCRP06} considering the transverse quasi-momentum $\vec{\kappa}_{\mb{\tiny T}} = k_{x}\widehat{e}_{x} + k_{y}\widehat{e}_{y}$, because this is the direction of the Brillouin Zone where is described the present KL Hamiltonian. We assume understood any modification of the selected Brillouin Zone direction, as a change in the model Hamiltonian to use. The system's quantal heterogeneity is considered along $z$ axis, taken perpendicular to the heterostructure interfaces [see Fig. \ref{F1-2Dn}(b)]. The operator $\widehat{\bn{W}}_{\mathrm{eff}}$, is nothing but somewhat arbitrary convention, valid as long as one get holds of all  {configuration functionals},  {such as} potential-like energy terms from the original Hamiltonian operator, which are $z$-component momentum free.\cite{Milanovic82,LCRP06} Then
\begin{equation}
  \label{for:Weff}
   \widehat{\bn{W}}_{\mathrm{eff}} =
    \left(
    \begin{array}{cccc}
      W_{11} & W_{12} & 0 & 0 \\
      W_{12}^{*} &  W_{22} & 0 & 0 \\
      0 & 0 &  W_{22} & W_{12} \\
      0 & 0 & W_{12}^{*} & W_{11}
     \end{array}
     \right)
  \end{equation}
\noindent is suitable for going through a standard calculation
\begin{equation}
  \label{eq:Veff}
  \left[\widehat{\bn{W}}_{\mathrm{eff}} - V_{\mathrm{eff}}\bn{I}_{4}\right]\bn{\Psi}(z) = \bn{O}_{4},
\end{equation}
\noindent leading us to the effective potential \emph{band offset} $V_{\mathrm{eff}}$, ``\emph{felt}" in some sense, by holes during their passage trough the heterostructure, as $\kappa_{\mb{\tiny T}}$ changes.

We introduced $W_{11(22)} = A_{1(2)}\kappa_{\mb{\tiny T}}^{2} + V(z)$ and $W_{12} = \frac{\hbar^{2}\sqrt{3}}{2\,m_{0}}\left(\gamma_{2}(k_{y}^{2}-k_{x}^{2}) + 2i\gamma_{3}k_{x}k_{y}\right)$, with $\gamma_{i}$ the L\"uttinger parameters, and $m_{0}$ the bare electron mass. The $\vec{\kappa}_{\mb{\tiny T}}$ components $k_{x,y}$, are set in-plane respect to the heterostructure interfaces. In (\ref{eq:Veff}) $\bn{\Psi}(z)$ is a multi-component envelope function. Though moderately rough, assertion (\ref{for:Weff}) represents a reliable-accuracy approximation to the $V_{\mathrm{eff}}$, {whose modifications }we are interesting in. Let us consider a periodic three-layer [$A$-cladding left ($L$) layer /$B$ middle ($M$) layer/ $A$-cladding right ($R$) layer] heterostructure, in the absence of external fields or strains. In the bulk cladding layers, \emph{hh} and \emph{lh} modes mix due to the $\bn{k}\cdot\bn{p}$ interaction, while the middle slab represents a inhospitable medium for holes.  At zero valence-band mixing, one has
\begin{eqnarray}
  \label{for:Veff-0}
  \hspace{-6mm}\bn{V}(z) =
  \left\{
   \begin{array}{lcl}
     0 &;& \,\, z < z_{\textsc l}\\
     V_{\textsc b} - V_{\textsc a} = V_{\mathrm{eff}} &;& z_{\textsc l} < z <z_{\textsc r}\\
     0 &;& \,\, z > z_{\textsc r}
   \end{array}
   \right\}
   = \Theta V_{\mathrm{eff}},
\end{eqnarray}
\noindent being $\Theta$ a step-like function, and $V_{\textsc {a/b}}$ the potential of the cladding/middle layer.  {Due the lack of strict superlattice multiple-layered structures under study, we have neglected the spontaneous in-layer polarization field for III-nitride constituent media \cite{Ambacher02}, thus assuming a rectangular potential profile as test-run input, rather than biased one for all envisioned III-nitride slabs of the heterostructures.}

 {Strain field may rise questions on their relative effects on the electronic structure and, in particular on the valence-band structure where shape and size of the potential profile lead to stronger hybridization of the quantum states.} Lets turn now to examine the effects of the stress, in the framework of the KL model Hamiltonian. The existence of a biaxial stress applied upon the plane parallel to the heterostructure interfaces, leads to the appearance of an in-plane strain.  The effective potential operator $\widehat{\bn{W}}_{\mathrm{eff}}$ (\ref{for:Weff}) in the presence of biaxial strain, can be written as \cite{Piprek03}
\begin{equation}
  \label{for:Weff-s}
   \widehat{\widetilde{\bn{W}}}_{\mathrm{eff}} = \widehat{\bn{W}}_{\mathrm{eff}} + U_{s}\bn{I}_{4},
\end{equation}
\noindent where
\begin{equation}
 \label{for:Strain}
  U_{s} = -\left\{a_{v}(2\varepsilon_{1} + \varepsilon_{3}) + b(\varepsilon_{1} - \varepsilon_{3}) \right\},
\end{equation}
\noindent is the accumulated strain energy resulting from the tensile or compressive stress acting on the crystal, when an epitaxial layer is grown on a different lattice-parameter substrate. Owing to strictness in formulation,\cite{Piprek03} we guess that a maximum-quota criterium (\ref{for:Strain}) it suffices to cover properly the aim posted in section \ref{sec:Funda}. So, being independent from $\kappa_{\mb{\tiny T}}$, a maximized $U_{s}$ was taken for granted, to evaluate if there is a real challenger strain effect respect to valence-band mixing influence  {on the metamorphosis of $V_{\mathrm{eff}}$}. Here, the subscript $s$ stands for strain. In (\ref{for:Strain}) $a_{v}$/$b$ represent the Pikus-Bir deformation/break potentials, describing the influence of hydrostatic/uniaxial strain. Meanwhile $\varepsilon_{1,3}$, are the in-plane, and normal-to-plane lattice displacements, respectively. For commonly used cubic and hexagonal semiconductor compounds, we assume\cite{RMohan10,Jovanovic05}
\begin{equation}
 \label{for:Epsi-plane}
  \varepsilon_{1} = -\frac{a_{s}-a_{l}}{a_{l}}
\end{equation}
\noindent being $a_{s,l}$ the lattice parameter of the substrate and the epitaxial layer, respectively. Though no external stress is considered along the growth direction $z$, the lattice parameter is forced to change due to the Poisson effect.\cite{Jovanovic05} Hence, the normal-to-plane displacement can be cast as
\begin{equation}
 \label{for:Epsi-normal}
 \varepsilon_{3} = -\nu\, \varepsilon_{1},
\end{equation}
\noindent which remains connected to in-plane deformation $\varepsilon_{1}$ via the Poisson radio $\nu$. The last is valid for zinc blende and wurtzite materials.

By changing the material and the growth plane, the value of $\nu$ modifies. For cubic materials it reads\cite{RMohan10}
\begin{equation}
 \label{for:Poison-c}
  \nu_{cub} = \left\lbrace
       \begin{array}{l}
         2\frac{C_{12}}{C_{11}} \qquad\qquad\qquad\textrm{for growth plane:}\:(001)\\
         \\
         \frac{C_{11} + 3C_{12} - 2C_{44}}{C_{11} + C_{12} + 2C_{44}} \qquad\textrm{for growth plane:}\:(110)\\
         \\
         \frac{2\left(C_{11} + 2C_{12} - 2C_{44}\right)}{C_{11} + 2C_{12} + 4C_{44}} \qquad\textrm{for growth plane:}\:(111)
       \end{array},\right.
\end{equation}
\noindent while for the hexagonal ones we have\cite{RMohan10}
\begin{equation}
 \label{for:Poison-h}
  \nu_{hex} = \left\lbrace
       \begin{array}{l}
         2\frac{C_{13}}{C_{33}} \qquad\qquad\qquad\textrm{for growth plane:}\:(0001)\\
         \\
         \frac{C_{12}\varepsilon_{1} + C_{13}\varepsilon_{c}}{C_{11}} \qquad\textrm{for growth plane:}\:(1\bar{1}00)\\
         \\
          \frac{C_{12}\varepsilon_{1} + C_{13}\varepsilon_{c}}{C_{11}} \qquad\textrm{for growth plane:}\:(1\bar{1}02)
       \end{array}.\right.
\end{equation}
\noindent To quote the parameter $\varepsilon_{c} = \left((c_{s} - c_{l})/c_{l}\right)$, we take  $c_{s}$ for the substrate wafer,\cite{Ram-Mohan01} while $c_{l}$ is referred to the epitaxially-grown layer on buffer stratum.

\section{Discussion of results}
\label{sec:Res&Dis}

\hspace{5mm}Unless otherwise specified, the graphical simulations of $V_{\mathrm{eff}}$ reported here, were calculated using highly specialized $III-V$ semiconductor binary(ternary)-compound cases for both unstressed and stressed cubic and hexagonal systems.  {The present numerical simulations consider different constituent media, regardless if they can be grown.} In this section, we briefly present numerical exercises within the \emph{root-locus-like} technique, to foretell  multiband-coupled charge-carrier effects for pseudomorphically stressed $III-V$ semiconductor layered systems.

\subsection{Simulation of $V_{\mathrm{eff}}$ profile evolution}
\label{sec:SimDis}

On general grounds, for $\kappa_{\mb{\tiny T}} \approx 0$ the $V_{\mathrm{eff}}$ \textbf{is constant} \cite{RF04,LCRP06}, while by letting grow $\kappa_{\mb{\tiny T}}$, the band mixing effects arise and $V_{\mathrm{eff}}$ changes \cite{Milanovic82,RF04,LCRP06}. Some authors had declared a shift upward in energy, of the boundstates in the effective potential well as the transverse wave vector increases \cite{Ekbote99}. We are focused here to evaluate first the stress-free systems, and then the effect of a pseudomorphic strain on $V_{\mathrm{eff}}$.

\subsubsection{Unstressed $V_{\mathrm{eff}}$ metamorphosis}
 \label{sec:SimDis-Free}

To gain some insight into the rather complicated influence of the band mixing parameter $\kappa_{\mb{\tiny T}}$, on the effective \emph{band offset}, we display several graphics in the present section. The central point here, is a reliable numerical simulation for the spatial distribution of $V_{\mathrm{eff}}$ while the valence-band mixing increases from $\kappa_{\mb{\tiny T}} \approx 0$ (uncoupled holes) to $\kappa_{\mb{\tiny T}} = 0.1$\AA$^{-1}$ (strong hole band mixing). This purpose requires a solution of (\ref{eq:Veff}) looking for a systematic start-point theoretical treatment of highly specialized $III-V$ semiconductor binary-compound cases of interest. We have set a width of $25\,$\AA$\;$ for the external cladding-layer $L$ and $R$, while for the middle one  we have taken a thickness of $50\,$\AA.

\begin{figure}
 \centering
 \subfigure[]{\includegraphics[width=75mm]{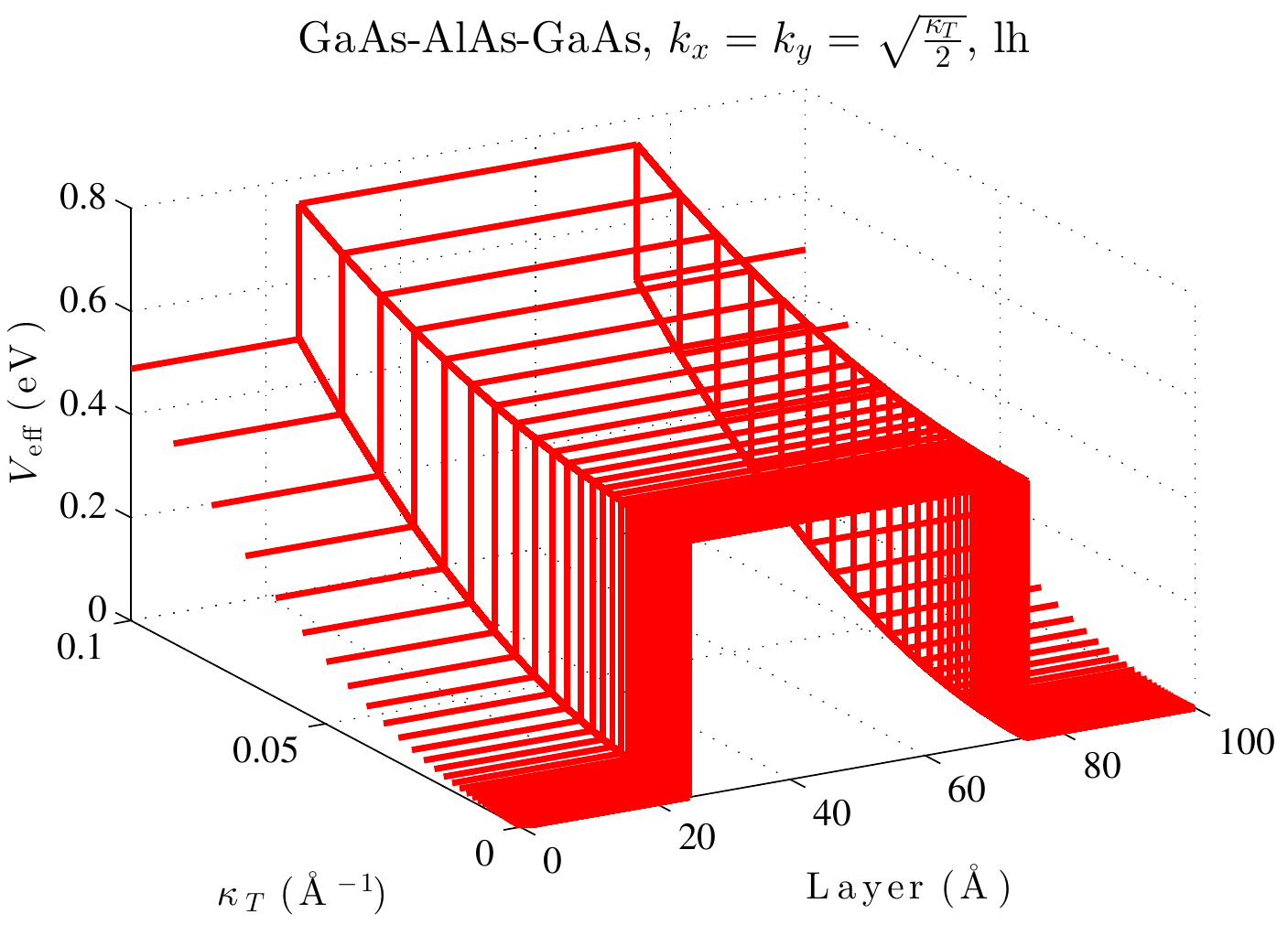}}
 \subfigure[]{\includegraphics[width=75mm]{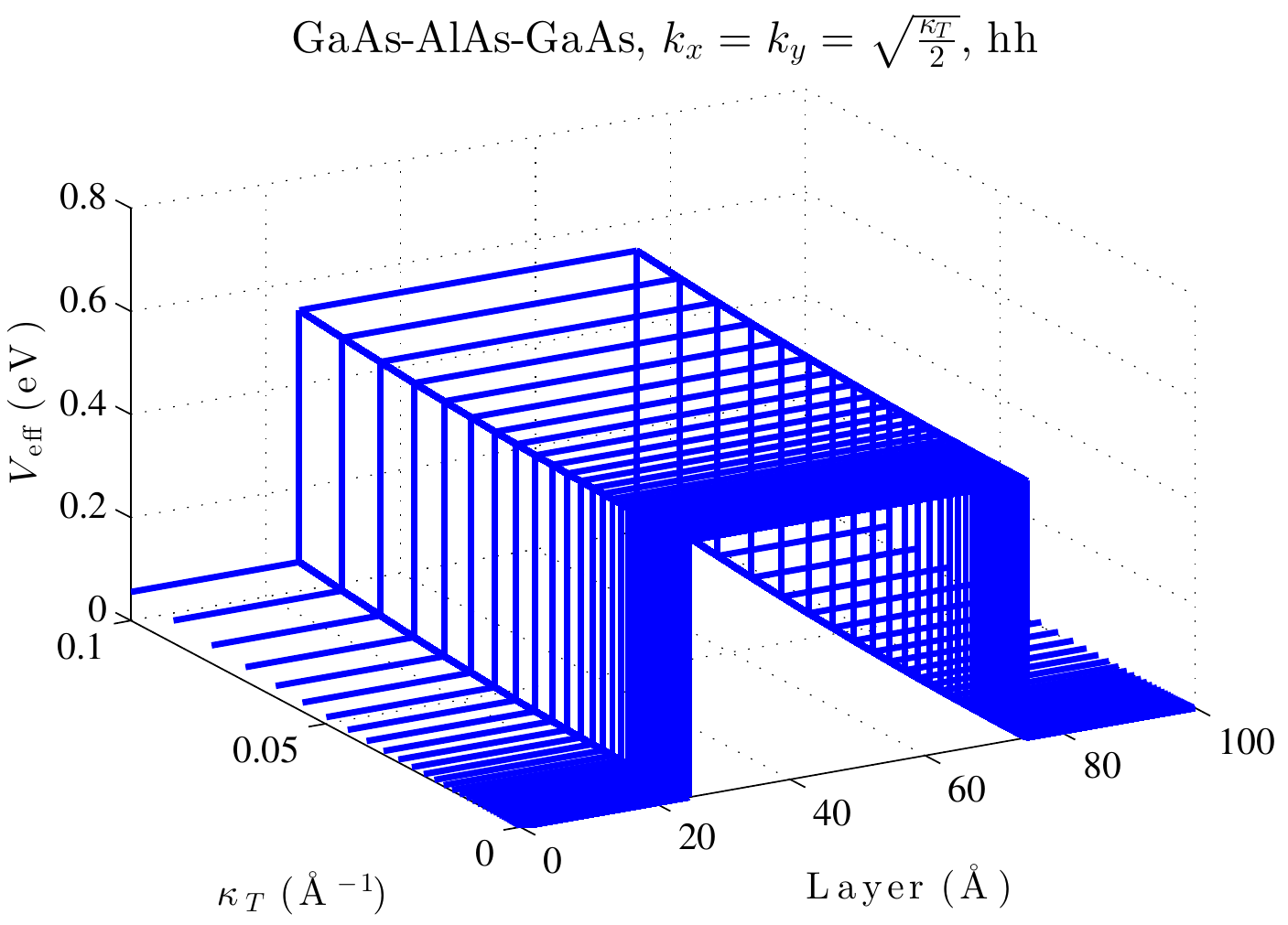}}
  \caption{\label{F2-3Dn} (Color online) Panel (a)/(b) displays the metamorphosis of the effective potential profile $V_{\mathrm{eff}}$ for \textit{lh}/\textit{hh} (red/blue lines), as a function of $\kappa_{\mb{\tiny T}}$  and layer dimension for a \textsl{GaAs}/\textsl{AlAs}/\textsl{GaAs} heterostructure.}
\end{figure}

Figure \ref{F2-3Dn} demonstrates that the standard  {fixed-height} rectangular distribution for $V_{z}$ (\ref{for:Veff-0}), is a consistent potential-energy trial  {of a QB} for \emph{hh} (blue lines), applicable in the wide range of $\kappa_{\mb{\tiny T}}$ [see panel (b)]. On the contrary, panel (a) remarks that the  {fixed-height QB}  is no longer valid for \emph{lh} (red lines), as $\kappa_{\mb{\tiny T}}$ increases. Is in this very sense, when the valence-band mixing effects get rise, that become unavoidable to refer an effective \emph{band offset} for a realistic description of the interplay of the envisioned physical structure with holes. We display in panel (a), the metamorphosis of $V_{\mathrm{eff}}$ for \emph{lh}, as a function of $\kappa_{\mb{\tiny T}}$  and layer dimension for a \textsl{GaAs}/\textsl{AlAs}/\textsl{GaAs} heterostructure. Two changes are neatly observable, namely: the energy edge of both left and right cladding-layers steps up in almost $0.5$ eV, while for the middle one it remains almost constant. The last departs from the $V_{\mathrm{eff}}$ evolution for \emph{hh}, where all borders move up almost rigidly [see panel (b)]. Although not shown here owing to brevity, a similar behavior was found for other middle-layer alloys (\textsl{AlSb}, \textsl{AlP}, \textsl{AlN}). Described above features, remain under modification of the in-plane direction.

An appealing situation arises, at a specific entry of the transverse momentum. An earlier detailed study on this subject, \cite{Milanovic82} had predicted the existence of such quantity $\kappa_{\mb{\tiny To}}^{2} = 2V_{o}m^{\textsc a}m^{\textsc b}\left[ \hbar^{2}\left( m^{\textsc b} - m^{\textsc a}\right)\right]$, for which $V_{\mathrm{eff}}$  becomes constant along the entire layered heterostructure. In the case envisioned here, due the presence of \emph{hh} and \emph{lh}, we have
\begin{equation}
 \label{for:Flat-Veff}
 \kappa_{\mb{\tiny To}(hh,lh)}^{2} = \frac{2V_{o}m_{o}^{3}\hbar^{2}}{\left(\gamma_{1}^{\textsc a} \mp 2\gamma_{2}^{\textsc a}\right)\left(\gamma_{1}^{\textsc b} \mp 2\gamma_{2}^{\textsc b}\right)} \left[\frac{1}{\left(\gamma_{1}^{\textsc b} \mp 2\gamma_{2}^{\textsc b}\right)} - \frac{1}{\left(\gamma_{1}^{\textsc a} \mp 2\gamma_{2}^{\textsc a}\right)} \right],
\end{equation}
\noindent being $V_{o} = V_{\mathrm{eff}}(\kappa_{\mb{\tiny T}} = 0)$, and $A/B$ standing for cladding/middle layer. A direct consequence for $V_{\mathrm{eff}}$ being flat at $\kappa_{\mb{\tiny To}}$ , is the existence of a crossover of $V_{\mathrm{eff}}$ respect to (\ref{for:Flat-Veff}). In other words, if a QW-like profile is present for $\kappa_{\mb{\tiny T}(hh,lh)}^{2} < \kappa_{\mb{\tiny To}(hh,lh)}^{2}$, then a QB-like profile appears at $\kappa_{\mb{\tiny T}(hh,lh)}^{2} > \kappa_{\mb{\tiny To}(hh,lh)}^{2}$, or the other way around.

\begin{figure}
\centering
  \subfigure[]{\includegraphics[width=2.83in]{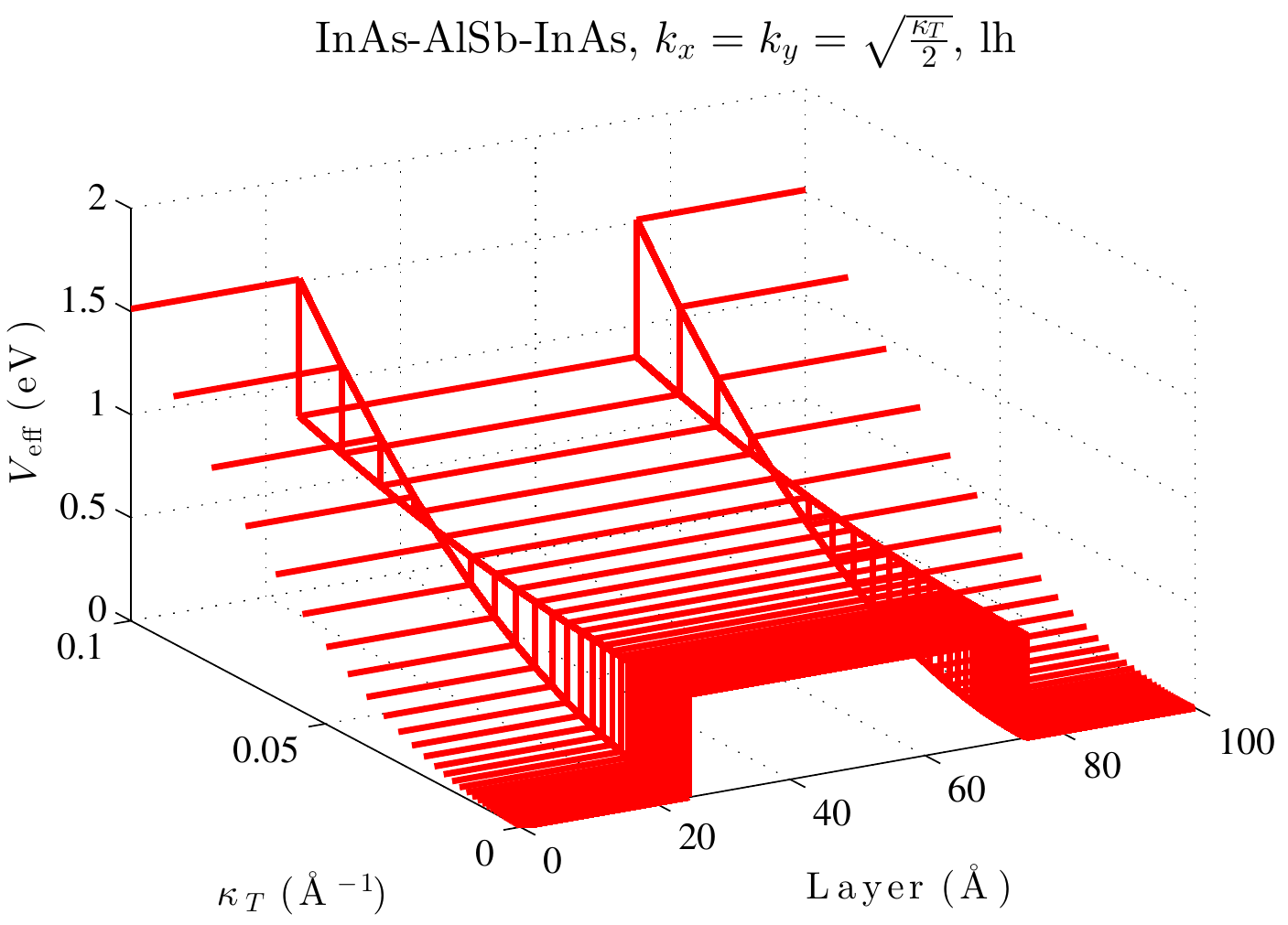}}
  \subfigure[]{\includegraphics[width=2.83in]{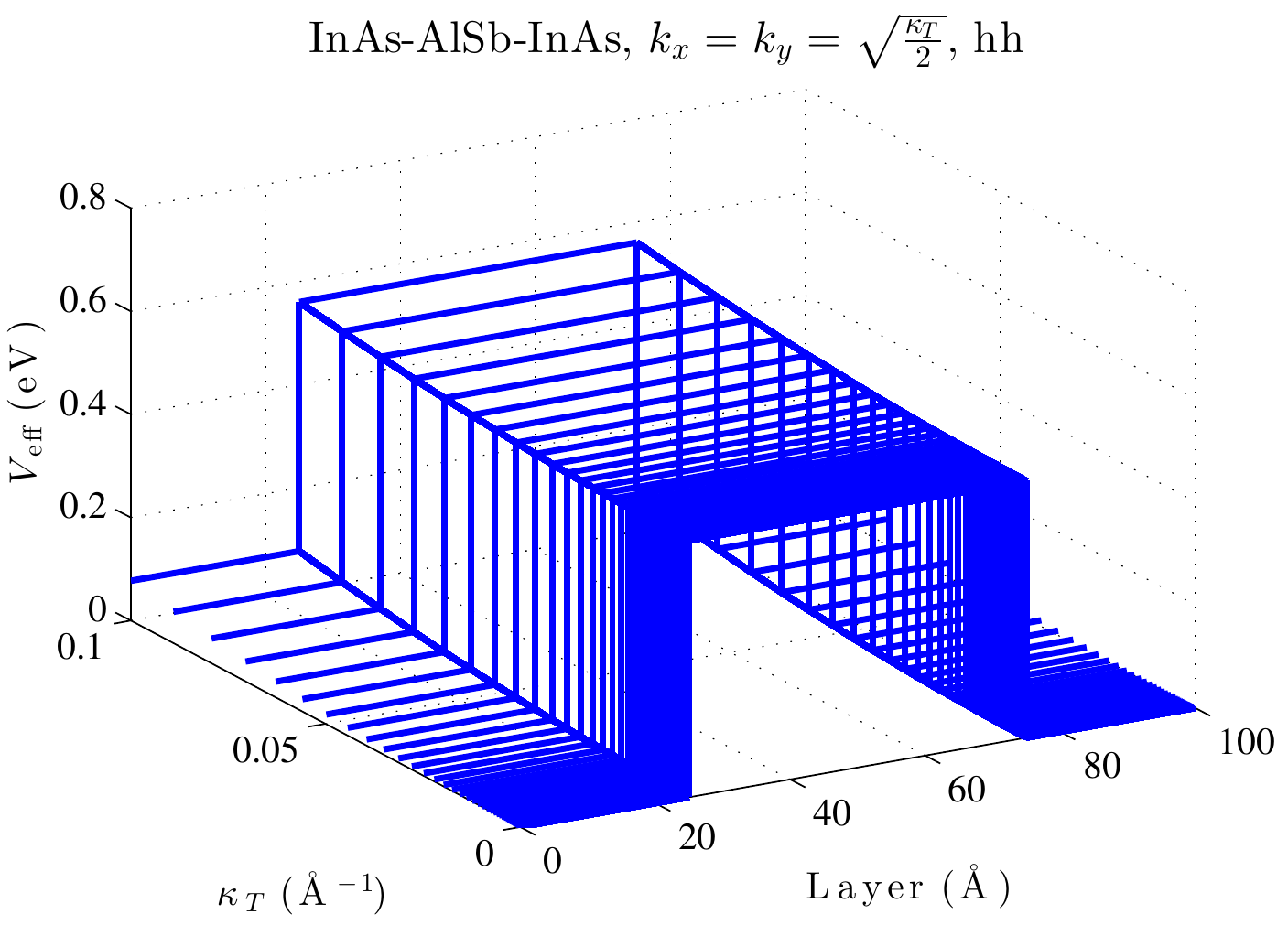}}\\
  \subfigure[]{\includegraphics[width=2.73in]{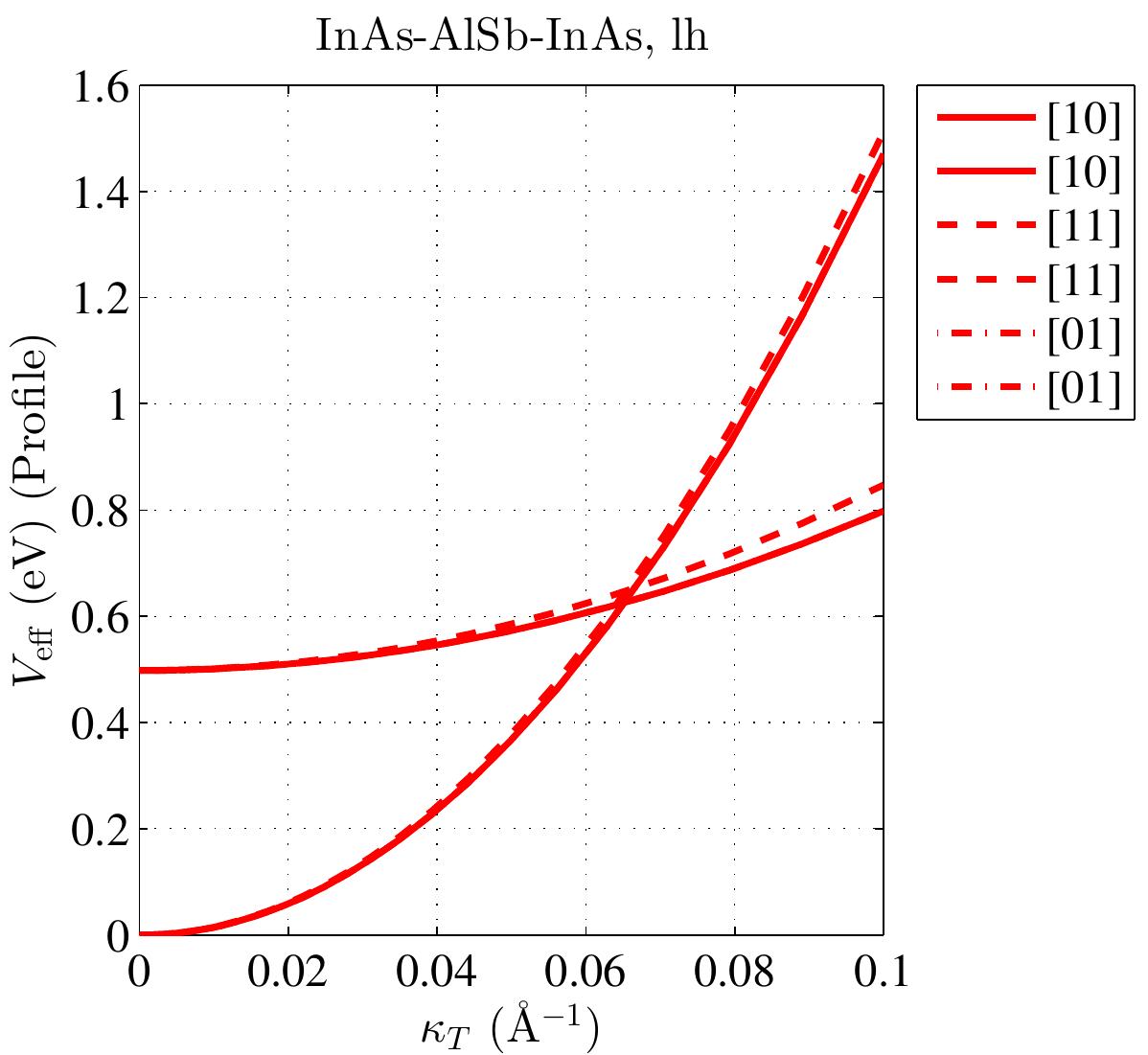}}
  \subfigure[]{\includegraphics[width=2.73in]{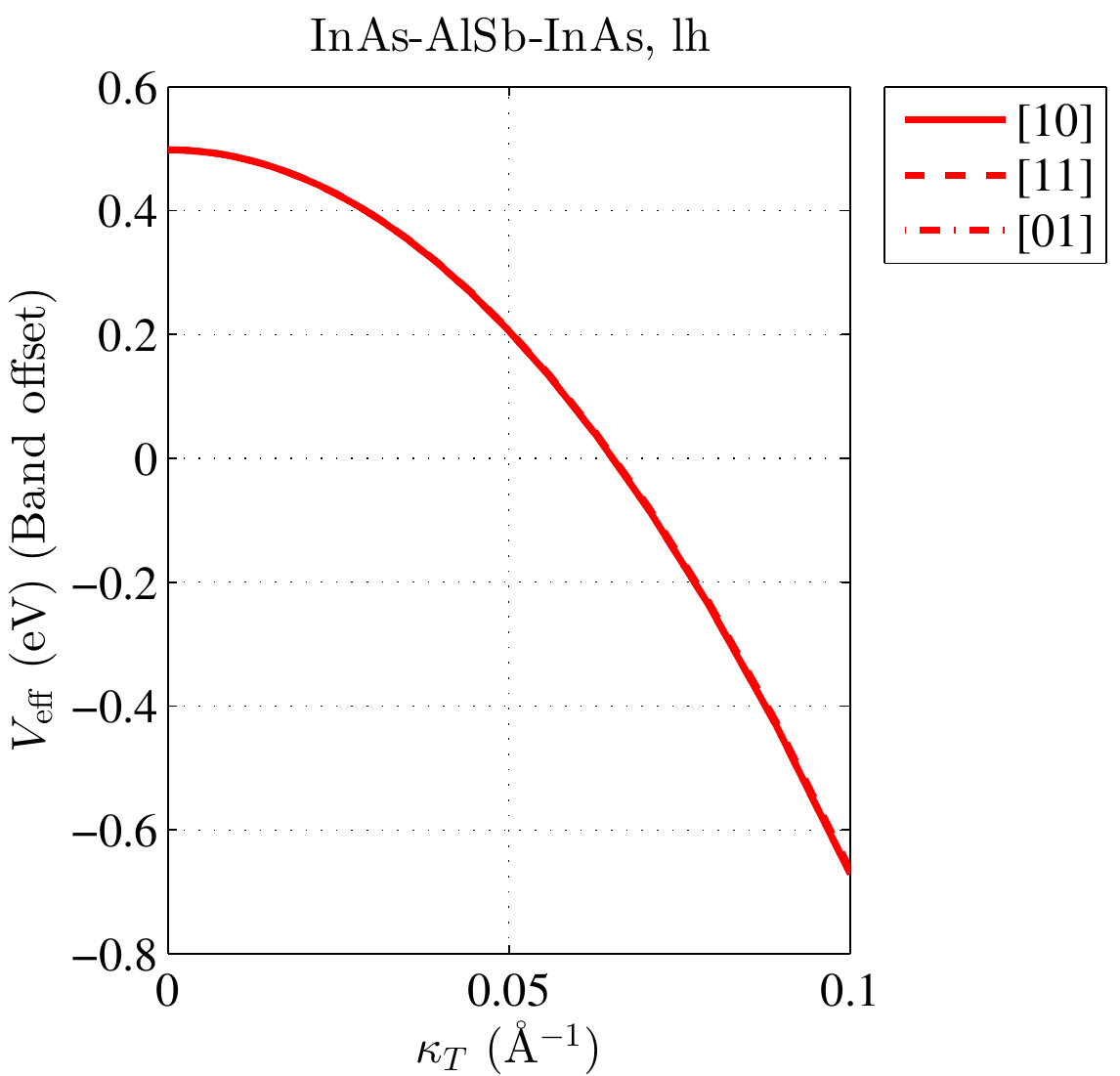}}
 \caption{\label{F6-3Dn} (Color online) Panel (a)/(b) displays the $3D$-perspective evolution of the $V_{\mathrm{eff}}$ profile for \emph{lh}/\emph{hh} (red/blue lines), as $\kappa_{\mb{\tiny T}}$  and layer dimension grow. Panel (c) displays a cut of the $V_{\mathrm{eff}}$ profile  for \emph{lh} (red line), at the interface plane between left and middle layers, as a function of $\kappa_{\mb{\tiny T}}$.  Panel (d) shows the progression of the \emph{band offset}, at the same interface for \emph{lh}, \textit{i.e.} the difference between the upper-edge and lower-edge of the $V_{\mathrm{eff}}$ profile. We have considered a \textsl{InAs}/\textsl{AlSb}/\textsl{InAs} stress-free layered heterostructure.}
\end{figure}

In Figure \ref{F6-3Dn}(a) the $V_{\mathrm{eff}}$ valence-band mixing dependence, exhibits a neatly {permutation} of  {the} $V_{\mathrm{eff}}$  {character} as the one predicted for electrons \cite{Milanovic82}. This  {permutation} pattern  {is what we call as} ``\emph{keyboard}" effect, {and} was detected for \emph{lh} only  {in stress-free systems}. This strike  {interchange of roles for QB-like} {and QW-like layers}, whenever the in-plane kinetic energy, varies from low to large intensity, represents the most striking contribution of the present study. For a single-band-electron Schr\"{o}dinger problem, some authors had predicted that both QW and QB may appears in the embedded layers of a semiconductor superlattice, depending on the transverse-component value of the wave vector.\cite{Milanovic82} Recently had been unambiguously demonstrated, that the effective-\emph{band offset} energy $V_{\mathrm{eff}}$, ``\emph{felt}'' by the two flavors of holes, as  $\kappa_{\mb{\tiny T}}$ grows, is not the same. Inspired on these earlier results, we had addressed a wider analysis of this appealing topic, displayed in Figure \ref{F6-3Dn}, pursuing a more detailed insight. We have considered a \textsl{InAs}/\textsl{AlSb}/\textsl{InAs} heterostructure. Panel (a)/(b) of Figure \ref{F6-3Dn} shows explicitly the metamorphosis of $V_{\mathrm{eff}}$, felt for both flavors of holes independently, respect to concomitant-material slabs. From panels (a) and (b), it is straightforward to see, that for \emph{hh} (blue lines), an almost constant $V_{\mathrm{eff}}$ remains, while $\kappa_{\mb{\tiny T}}$ varies from $0$ (uncoupled holes) to $0.1$\AA$^{-1}$ (strong hole band mixing), despite the respective band-edge levels had changed. At variance with the opposite for \textit{hh} [blue lines, panel (b)], the \emph{lh} exclusively [red lines, panel (a)] exhibit the  {\emph{keyboard} effect},  {\textit{i. e.} they} feel an effective \emph{band offset} exchanging from a QW-like into a QB-like one, and viceversa for an \textsl{InAs}/\textsl{AlSb}/\textsl{InAs} heterostructure, while $\kappa_{\mb{\tiny T}}$ increases. The {evident \emph{keyboard} effect} of $V_{\mathrm{eff}}$, resembles a former prediction for electrons \cite{Milanovic82}. This observation means, that in the selected rank of parameters for a given binary-compound materials, a \textit{lh} might ``\emph{felt}'' a qualitative different $V_{\mathrm{eff}}$ (QW or QB), during its passage through a layered system, while it is varying the degree of freedom in the transverse plane. Former assertions can be readily observed in Figure \ref{F6-3Dn}(c)-(d), were we had plotted the evolution of $V_{\mathrm{eff}}$ profile [panel (c)], as well as  the progression of the \emph{band offset} [panel (d)], with $\kappa_{\mb{\tiny T}}$ at a fixed transverse plane of the heterostructure. Both upper-edge and lower-edge move in opposite directions [see panel (c)] and the zero-\emph{band offset} point configuration is detected in the vicinity of $\kappa_{\mb{\tiny T}} \approx 0.066$\AA$^{-1}$ [see panel (d)]. The  {permutation} holds for other in-plane directions, as can be seen from panel (c). Although not shown here for simplicity, the \emph{keyboard} effect, remains robust for other middle-layer binary compounds, namely: \textsl{AlAs}, \textsl{AlP}, and \textsl{AlN}.

\subsubsection{{Keyboard} effect \texttt{\textit{versus}} pseudomorphic strain.}
 \label{sec:SimDis-Stress}

Turning now to built-in elastic stressed layered heterostructures, we are interested to answer a simple question: wether or not the existence of a pseudomorphic strain becomes  {a weak or a strong} competitor mechanism, able  {sometimes just} to diminish the \emph{keyboard} effect on $V_{\mathrm{eff}}$, or even {make it rises/vanishes} occasionally. Thereby, we need to account the accumulated strain energy resulting from the tensile or compressive stress acting on the  crystal slabs. The last requires to solve (\ref{for:Weff-s}), presuming the heterostructure sandwiched into a pseudomorphically strained QW/QB/QW-sequence.

\begin{figure}
\centering
  \subfigure[]{\includegraphics[width=2.83in]{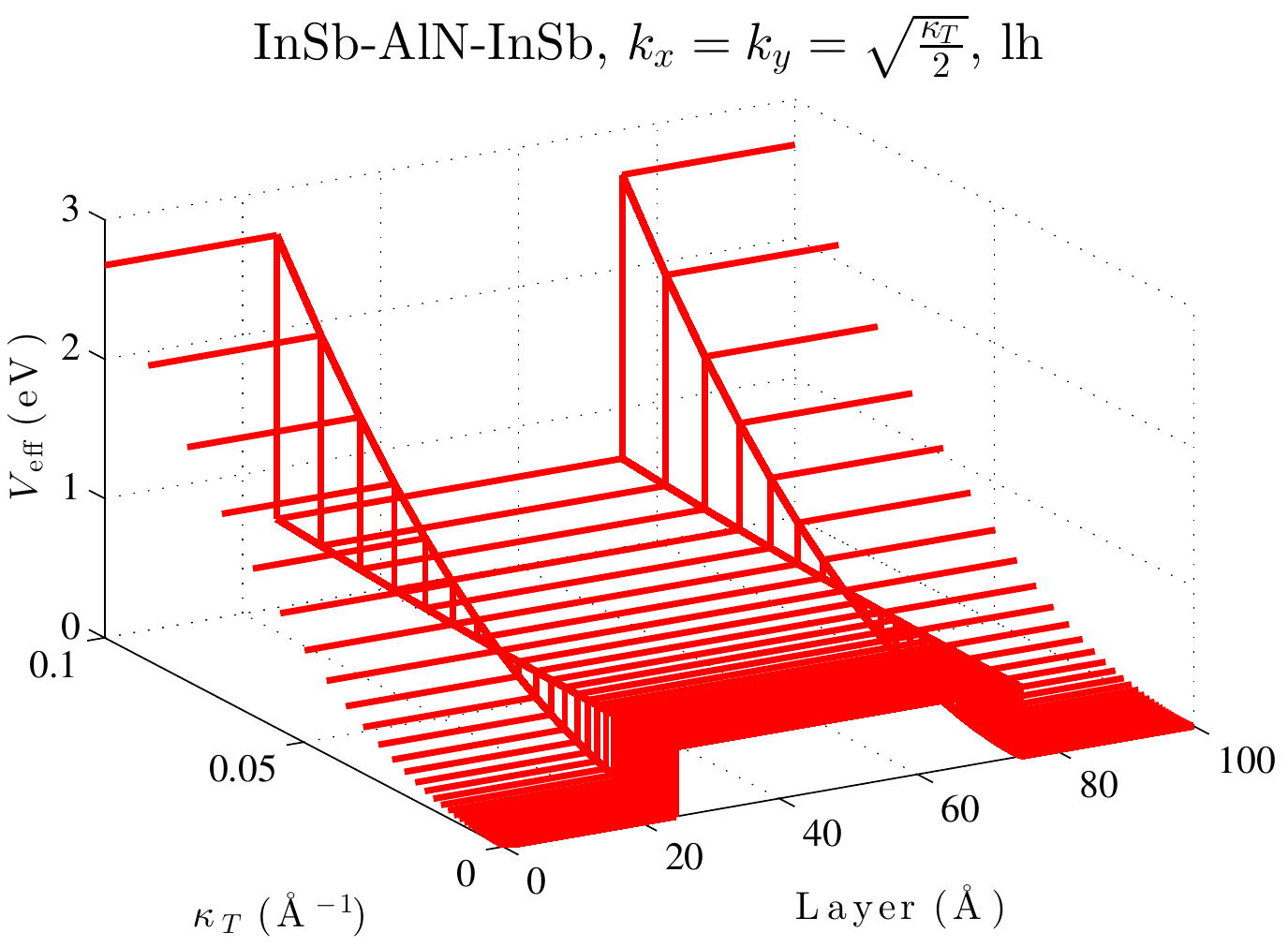}}
  \subfigure[]{\includegraphics[width=2.83in]{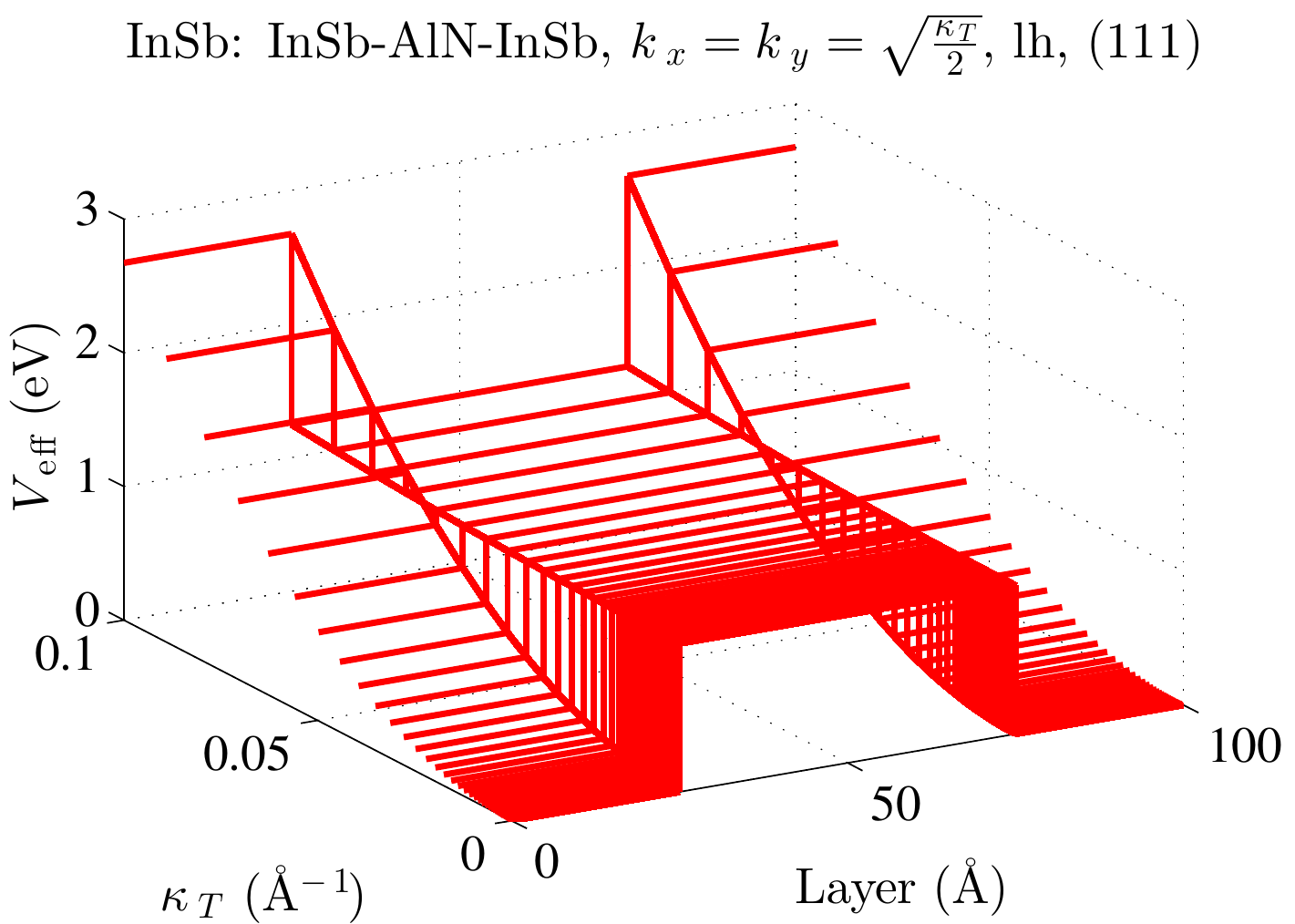}}
 \caption{\label{F7-3D} (Color online) Panel (a) displays the $3D$-perspective evolution of the stress-free $V_{\mathrm{eff}}$ profile for \emph{lh} as $\kappa_{\mb{\tiny T}}$  and layer dimension grow. Panel (b) shows the same for a \textsl{InSb}:\textsl{InSb}/\textsl{AlN}/\textsl{InSb} strained layered heterostructure.}
\end{figure}

Figure \ref{F7-3D} is devoted to demonstrate that the \emph{keyboard} pattern for \emph{lh} remains robust in a \textsl{InSb}:\textsl{InSb}/\textsl{AlN}/\textsl{InSb} pseudomorphically strained layered heterostructure [see panel (b)], respect to that of the stress-free system [see panel (a)]. In this case, we conclude that maximized $U_{s}$ (\ref{for:Strain}) do not represent any antagonist  {mechanism} regarding to valence-band mixing influence on $V_{\mathrm{eff}}$. 

\subsection{Influence of the pseudomorphic strain on $k_{z}$-spectrum}

 The QEP $k_{z}$-spectrum  {is a meaningful},  {and well-founded physical quantity}  {that can be} obtained \textit{via }the \emph{root-locus-like} procedure \cite{AJGL11} by unfolding back in the complex plane the dispersion-curve values for bulk materials, determined by stress-induced effects on the stress-free heterostructure.  {Thus, we take advantage of the \emph{root-locus-like} know-how, to promptly}  {identify evanescent modes, keeping in mind that complex (or pure imaginary) solutions are forbidden}  {for some layers and represent unstable solutions underlining the lack of hospitality of these slabs for oscillating modes}.  {The opposite examination is straightforward and also suitable for propagating modes, which become equated with stable solutions for given layers}.


 {To obtain the QEP $k_{z}$-spectrum in a periodic pseudomorphically}  {strained heterostructures of QB-acting/QW-acting/QB-acting materials, we first use (\ref{for:Weff-s}) and substitute it in (\ref{for:Weff}). Next, we solve again the characteristic problem (\ref{eq:Veff}), whose eigenvalues allow us to obtain the new expression for the QEP-matrix $\bn{\mathcal{K}}$, and then finally consequently solve (\ref{eq:QEP}) for $k_z$.} Once we have quoted the eigenvalues $k_{z}$ of (\ref{eq:Polinom}), it is then straightforward to generate a plot in the complex plane, symbolizing the {locations} of $k_{z}$ values that rise as a band mixing parameter $\kappa_{\mb{\tiny T}}$ changes. Keeping in mind that complex (or pure imaginary)/real solutions of (\ref{eq:Polinom}) represent forbidden/allowed modes, we take advantage of the \emph{root-locus-like} map to identify evanescent/propagating modes for a given layer. Thus, we are able ``to stamp'' on a $2D$-map language, a frequency-domain analysis of the envisioned heterostructure under a quantum-transport problem. This way, we are presenting an unfamiliar methodology in the context of quantum solid state physics, to deal with low-dimensional physical phenomenology.

\begin{figure}
\subfigure[]{\includegraphics[width=3in]{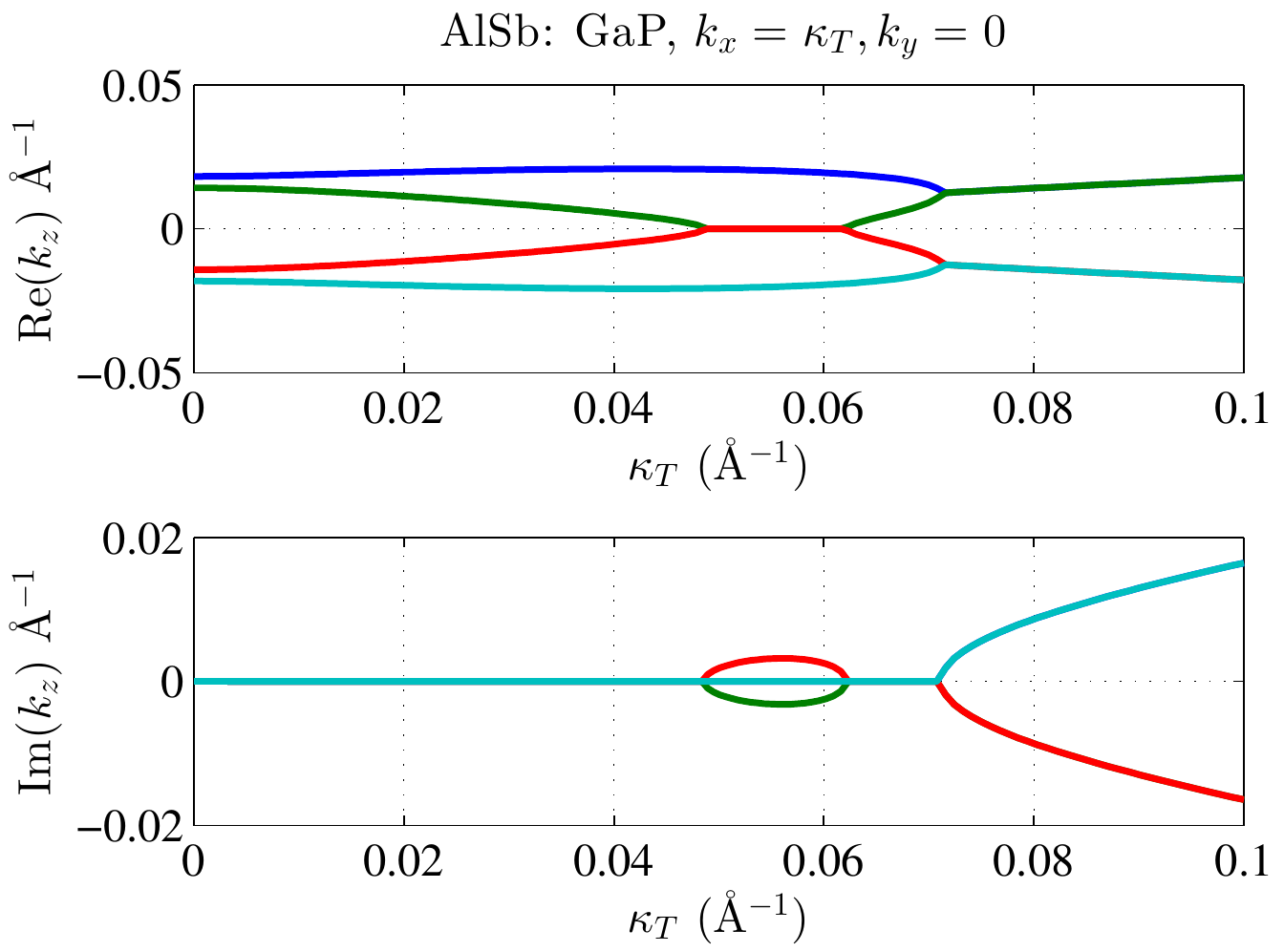}}
\subfigure[]{\includegraphics[width=3in]{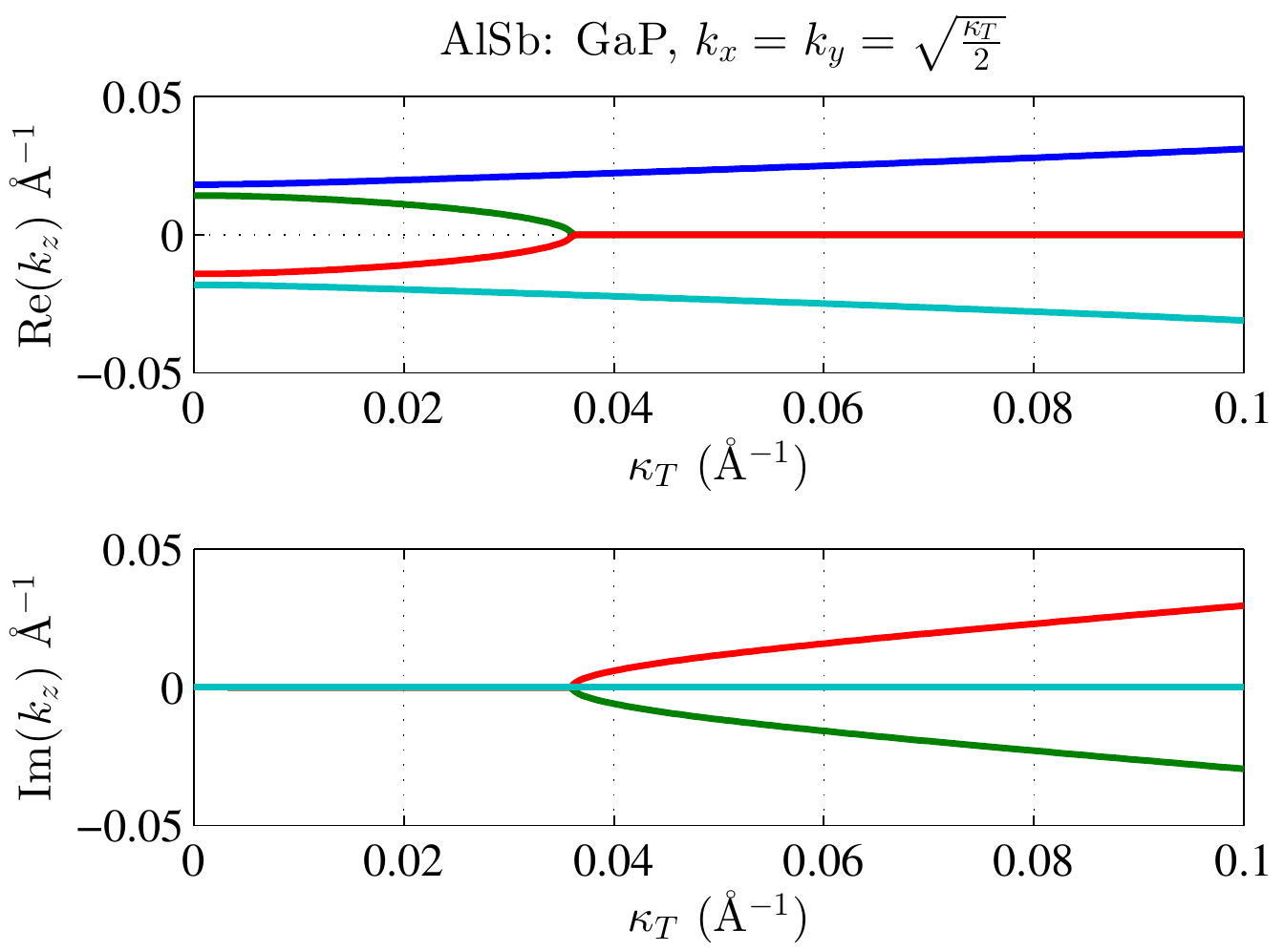}}
\caption{\label{F13-2Dn}(Color online) Root locus for the eigenvalues $k_{z}$ from QEP (\ref{eq:Polinom}), as a function of $\kappa_{\mb{\tiny T}}$ for strained \textsl{AlSb}(substrate)/\textsl{GaP} (epitaxial layer). We had assumed $E=0.6$ eV, and in-plane directions [10]/[01] for panel (a)/(b).}
\end{figure}

\begin{figure}
\subfigure[]{\includegraphics[width=3in]{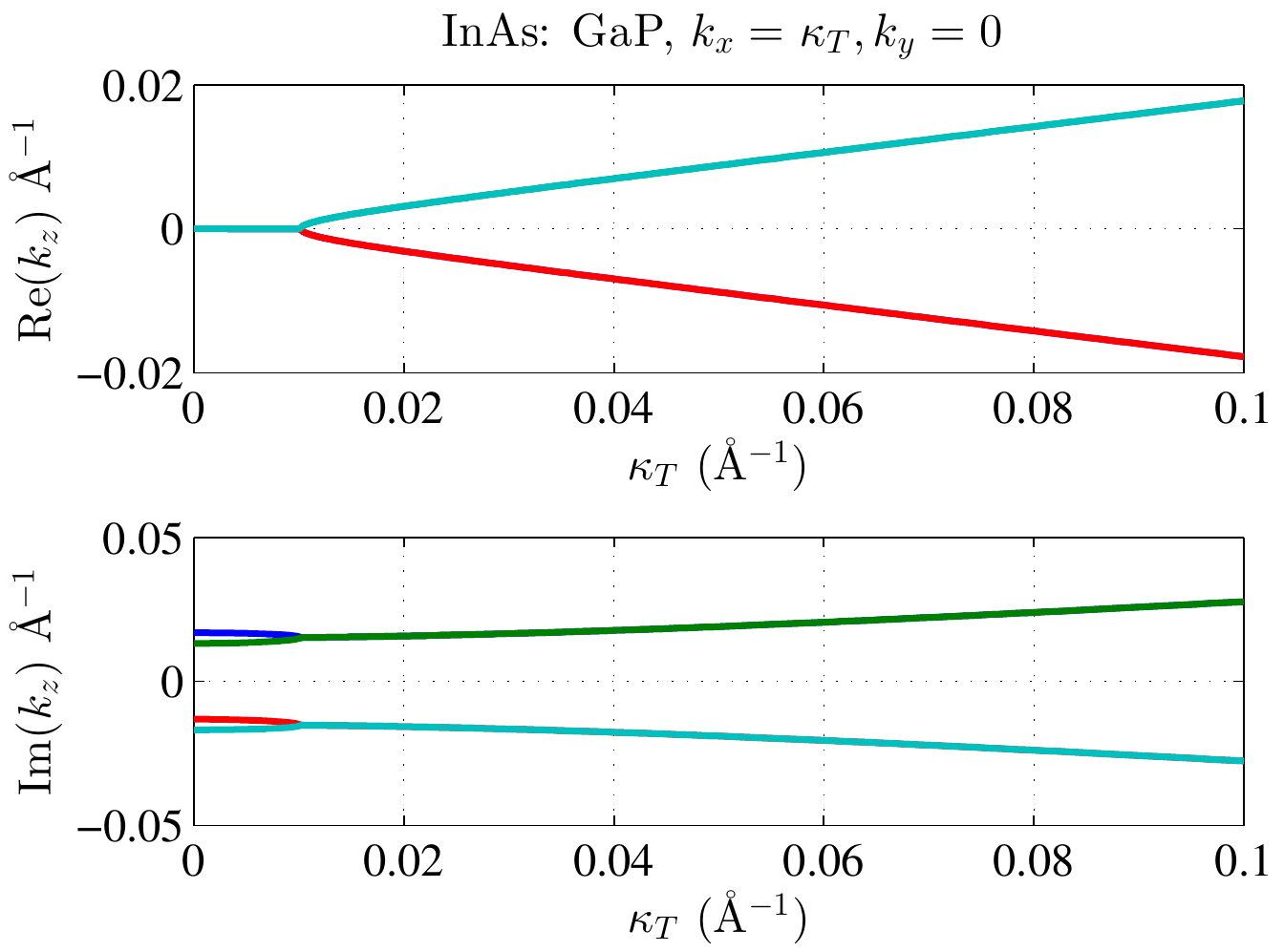}}
\subfigure[]{\includegraphics[width=3in]{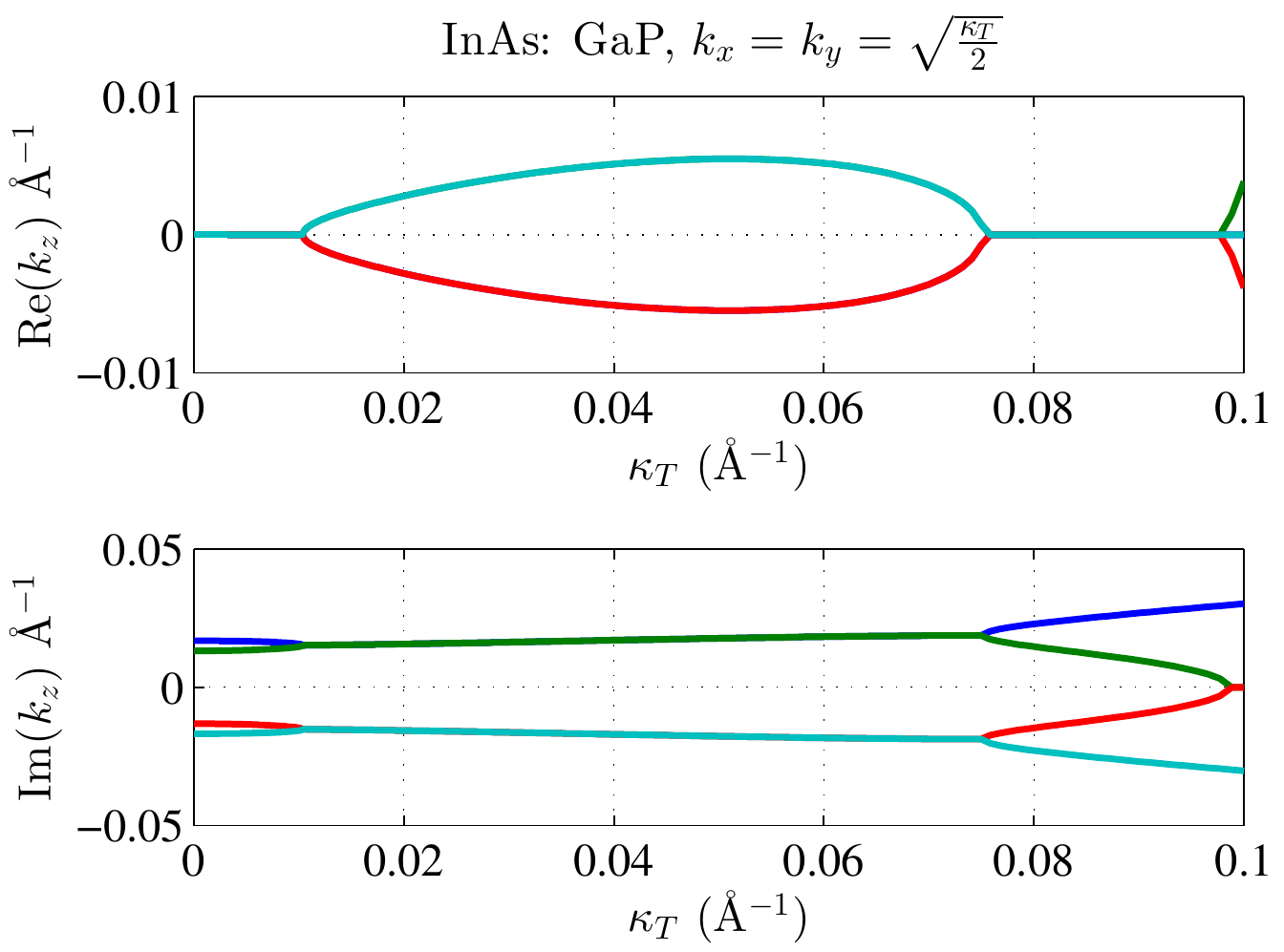}}
\caption{\label{F14-2Dn}(Color online) Root locus for the eigenvalues $k_{z}$ from QEP (\ref{eq:Polinom}), as a function of $\kappa_{\mb{\tiny T}}$ for strained \textsl{InAs}(substrate)/\textsl{GaP} (epitaxial layer). We had assumed $E=0.45$ eV, , and in-plane directions [10]/[01] for panel (a)/(b).}
\end{figure}

The Figure \ref{F13-2Dn} and Figure \ref{F14-2Dn}, illustrate the role of band mixing for $\kappa_{\mb{\tiny T}} \left[10^{-6}, 10^{-1}\right]$ \AA$^{-1}$, on the $k_z$ spectrum from QEP (\ref{eq:Polinom}), for a $III-V$ strained alloy, clearly distinguished as QW in most layered systems with technological interest. Importantly, by assuming two different substrates $AlSb$ (Fig.\ref{F13-2Dn}) and $InAs$ (Fig.\ref{F14-2Dn}), we found different patterns of the $k_{z}$ spectrum for $lh$ and $hh$. Namely for the $[10]$ in-plane direction, the  $k_{z}$ \emph{root-locus-like} evolution is real for $lh$, in the range of $\kappa_{\mb{\tiny T}} \in \left[10^{-6},0.049 \right]$ \AA$^{-1}$ and $\kappa_{\mb{\tiny T}} \in \left[0623, 0.0708\right]$ \AA$^{-1}$, while in the intervals $\kappa_{\mb{\tiny T}} \in \left[0.049, 0.0623\right]$ \AA$^{-1}$ and $\kappa_{\mb{\tiny T}} \in \left[0.0708, 0.1\right]$ \AA$^{-1}$, $k_{z}$ becomes pure imaginary and complex, respectively [see Fig.\ref{F13-2Dn}(a), inner green-red solid lines]. On the other hand, the  $k_{z}$ \emph{root-locus-like} shows real values for $hh$, in the interval $\kappa_{\mb{\tiny T}} \in \left[10^{-6}, 0.0708\right]$ \AA$^{-1}$ and is complex, when $\kappa_{\mb{\tiny T}}\in \left[0.0708, 0.1\right]$ \AA$^{-1}$ [see Fig.\ref{F13-2Dn}(a) outer blue solid lines]. Worthwhile stress that $hh$ and $lh$ curves, are undistinguishable in this last interval. Although not shown here, the $[01]$ in-plane direction exhibits the same behavior. The Fig.\ref{F13-2Dn}(b), displays the QEP (\ref{eq:Polinom} spectrum along the $[11]$ in-plane direction. For $lh$ only, $k_{z}$ \emph{root-locus-like} evolution starts as a real number in the range $\kappa_{\mb{\tiny T}} \in \left[10^{-6}, 0.0363\right]$ \AA$^{-1}$, and becomes pure imaginary for $\kappa_{\mb{\tiny T}} \in \left[0.0363, 0.1\right]$ \AA$^{-1}$. The $k_{z}$ spectrum for $hh$ it is always a real number in the whole selected interval $\kappa_{\mb{\tiny T}} \in \left[10^{-6}, 0.1\right]$ \AA$^{-1}$. Panel (a) of Fig.\ref{F14-2Dn} describes in-plane direction $[10]$, and for analogy the $[01]$ --although not depicted for brevity--, with the band mixing. For both $lh$ and $hh$, the $k_{z}$ evolution starts as a pure imaginary number  in the range $\kappa_{\textsc t} \in \left[10^{-6}, 0.01\right]$ \AA$^{-1}$, and become a complex number in the interval $\kappa_{\mb{\tiny T}} \in \left[0.01, 0.1\right]$ \AA$^{-1}$. In this gap the $hh$ and $lh$ are indistinguishable, as their $k_{z}$ magnitude is the same. Meanwhile, the panel (b) of Fig.\ref{F14-2Dn} demonstrates that for the $[11]$ in-plane direction, the $k_{z}$ values are mostly complex or pure imaginary, except in the small interval of $\kappa_{\mb{\tiny T}} \in \left[0.097,0.1\right]$ \AA$^{-1}$, where they are real. None real entries of $k_{z}$ for $hh$, were found as $\kappa_{\mb{\tiny T}}$ changes within the bounds $\left[0.01, 0.1\right]$ \AA$^{-1}$. The $hh$ and $lh$ curves are indistinguishable in the range of $\kappa_{\mb{\tiny T}} \in \left[0.0133,0.075\right]$ \AA$^{-1}$. After this detailed description, several features deserve close attention. In short: the $[10]$ and $[01]$ in-plane directions, show an isotropic behavior, for each selected substrate. The real-value domains of the \emph{root-locus-like} map of $k_{z}$, means that the $GaP$ strained-layer recovers his standard QW-behavior for both $hh$ and $lh$ quasi-particles, regarding the stress-free configuration. On the opposite, whenever real-value map fades, \textit{i.e.} complex or pure imaginary magnitudes arise, none oscillating modes can propagate through an $InAs:GaP$ strained slabs. In this last case, the $GaP$ might turns into an effective QB, for traveling holes.


\section{\label{sec:Conclu}Conclusions}

We present an alternative procedure to simulate graphically, the phenomenon of the transverse degree of freedom  {influence on} the effective scattering potential. For low-intensity valence-band mixing regime, a  {fixed-height} rectangular distribution of the potential-energy, is a good trial as a standard reference frame, for a theoretical treatment involving both flavors of holes under study. However this assertion is no longer valid, whenever the mixing effects reveal. At variance with the opposite for \emph{hh}, the \emph{lh} exclusively,  {experience the strike \emph{keyboard} effect and permutations of $V_{\mathrm{eff}}$}  {in stress-free systems}.  
 Our results provide an unambiguous demonstration for the apparent robustness of the  {fixed-height} flat $V_{\mathrm{eff}}$ as test run input whenever pure $hh$ and $lh$, are mixed. Pseudomorphic strain is able to diminish the \emph{keyboard} effect on $V_{\mathrm{eff}}$, and also makes it emerge or even vanish eventually. We conclude that the multiband-mixing effects modulated by stress induced events,  {are competitors mechanisms} that can not be universally neglected by assuming a { fixed-height rectangular spatial} distribution for {fixed-character} potential energy,  {as a reliable test-run input for semiconducting heterostructures}. {Present modelling of $V_{\mathrm{eff}}$ evolution, may be a reliable workbench for testing other configurations, besides our results may be of relevance for promising heterostructure's design guided by valence-band structure modeling to enhance the hole mobility in III-V semiconducting materials provided they always lagged compared to II-IV media \cite{Nainani11}.}

\section*{Acknowledgments}
This work was developed under support of DINV, UIA, M\'{e}xico. One of the authors (L.D-C) is grateful to the Visiting Academic Program of the UIA, M\'{e}xico.


\section*{References}
\begin{thebibliography}{20}
\bibitem{Klicmeck01} G. Klicmeck, R. Ch. Bowen, and T. B. Boykin, Superlattices and Microstructures \textbf{29}, 187 (2001).
\bibitem{Wessel89}R. Wessel and M. Altarelli, Phys. Rev. B 39, 12802 (1989).
\bibitem{Schneider89} H. Schneider, H. T. Grahn, K. Klitzing, and K. Ploog, Phys. Rev. B \textbf{40}, 10040 (1982).
\bibitem{Bittencourt97} A. C. Bittencourt, A. M. Cohen and G. E. Marques, \textit{Brazilian J. Phys.} \textbf{27}, 281 (1997).
\bibitem{Foreman07} Bradley A. Foreman, \textit{Phys. Rev. B} \textbf{76}, 045327 (2007)
\bibitem{Ekins99} N. J. Ekins-Daukes, K. W. J. Barnham, J. P. Connolly, J. S. Roberts, J. C. Clark, G. Hill and M. Mazzer, \textit{App. Phys. Lett.} \textbf{75}, 4195 (1999).
\bibitem{Smeeton03} T. M. Smeeton, M. J. Kappers, J. S. Barnard, M. E. Vickers and C. J. Humphreys, \textit{App. Phys. Lett.} \textbf{83}, 5419 (2003).
\bibitem{Nainani11} Aneesh Nainani, Brian R. Bennett, J. Brad Boos, Mario G. Ancona and Krishna C. Saraswat, arxiv.org/pdf/1108.5507 (2011).
\bibitem{Faux97} D. A. Faux, J. R. Downes and E. P. OReilly, \textit{J. App. Phys.} \textbf{82}, 3754 (1997).
\bibitem{Sunil08} Sunil Patil, W. P. Hong and S. H. Park, \textit{Phys. Lett. A} \textbf{372}, 4076 (2008).
\bibitem{Bafna13} Manish K. Bashna, Pratima Sen and P. K. Sen, \textit{Indian J. Pure App. Phys.} \textbf{51}, 553 (2013).
\bibitem{Milanovic82} V. Milanovic, and D. Tjapkin, \textit{Phys. Stat. Sol(b)} \textbf{110}, 687 (1982).
\bibitem{RF04} Rolando P\'erez-\'Alvarez and Federico Garc\'{\i}a-Moliner,``\emph{Transfer Matrix, Green Function and related techniques:Tools for the study of multilayer heterostructures}'', (Ed. Universitat Jaume I, Castell\'{o}n de la Plana, Espa\~{n}a), 2004.
\bibitem{Ekbote99} S. Ekbote, M. Cahay and K. Roenker, \textit{J. App. Phys.} \textbf{85}, 924 (1999).
\bibitem{LCRP06} L. Diago-Cisneros, H. Rodr\'{\i}guez-Coppola, R. P\'erez-\'Alvarez, and P. Pereyra, \textit{Phys. Rev. B} \textbf{74}, 045308 (2006).
\bibitem{RMohan10} K. H. Yoo, J. D. Albrecht and L. R. Ram-Mohan, \textit{Am. J. Phys.}, \textbf{78}, 589 (2010).
\bibitem{Jovanovic05} V. D. Jovanovi\'{c}, ``\emph{Quantum Wells, Wires and Dots}", (Ed. John Wiley \verb"&" Sons, Ltd.), 2005.
\bibitem{Piprek03} Joachim Piprek, ``\emph{Semiconductor Optoelectronic Devices. Introduction to Physics and simulation}", (Ed. Elesevier, Acadenic Press), 2003.
\bibitem{AJGL11} A. Mendoza-\'{A}lvarez, J. J. Flores-Godoy, G. Fern\'{a}ndez-Anaya,  and L. Diago-Cisneros, \textit{Phys. Scr.} \textbf{84}, 055702 (2011).
\bibitem{JALG13} J. J. Flores-Godoy, A. Mendoza-\'{A}lvarez, L. Diago-Cisneros, and G. Fern\'{a}ndez-Anaya \textit{Phys. Status Slidi B}, \textbf{67}, 1339 (2113).
\bibitem{Ambacher02} O. Ambacher, J. Majewski, C. Miskys, A. Link, M. Hermann, M. Eickhoff, M. Stutzmann, F. Bernardini, V. Fiorentini, V. Tilak, B. Schaff, and L. F. Eastman, \textit{J. Phys.: Condens. Matter} \textbf{14}, 3399 (2002).
\bibitem{Ram-Mohan01} I. Vurgaffman, J. R. Mayer, and L. R. Ram-Moham, \textit{J. App. Phys.} \textbf{89}, 5815 (2001).


\end{thebibliography}
\end{document}